\documentclass{elsart}

\usepackage{natbib}

\usepackage{epsfig}

\usepackage{amssymb}

\begin{document}

\begin{frontmatter}


%

\title{Using nonlocal electrostatics for solvation free energy computations:
ions and small molecules}

\author{A. Hildebrandt}\ead{anhi@bioinf.uni-sb.de, Fax: +49 681 302 64719}
\author{, O. Kohlbacher, R. Blossey, H.-P. Lenhof}
\address{Center for Bioinformatics,
Saarland University, Building 36.1, P.O.Box 151150, 66041
Saarbr\"ucken, Germany}

\begin{abstract}
Solvation free energy is an important quantity in Computational Chemistry with
a variety of applications, especially in drug discovery and design. The
accurate prediction of solvation free energies of small molecules in water is
still a largely unsolved problem, which is mainly due to the complex nature of
the water-solute interactions. In this letter we develop a scheme for the
determination of the electrostatic contribution to the solvation free energy
of charged molecules based on nonlocal electrostatics involving a minimal
parameter set which in particular allows to introduce atomic radii in a
consistent way. We test our approach on simple ions and small molecules for
which both experimental results and other theoretical descriptions are
available for quantitative comparison. We conclude that our approach is both
physically transparent and quantitatively reliable.
\end{abstract}

\begin{keyword}
solvation free energy \sep nonlocal electrostatics
\end{keyword}

\end{frontmatter}

\section{Introduction}
\label{1}

The unique properties of water, namely its polar nature, high dielectric
constant, and its ability to form hydrogen bonds \cite{ball}, are responsible
for the existence of life as we know it. At the same time, these very
properties are the main obstacle in modelling hydration effects.  An accurate
model of ion and small molecule hydration will therefore have important applications in
computational chemistry, chemical engineering, and drug design.

In the latter, the interest stems from the fact that drug-like small molecules
are usually moderately polar to achieve the required bioavailability.
Therefore their solvation free energy is generally dominated by the
electrostatic contribution, while the nonpolar contribution is more or less
negligible. One of the most successful models for predicting solvation free
energies of small molecules (and proteins) therefore employs the Poisson or
Poisson-Boltzmann equations \cite{hon93,hon95}. Within this approach, the
solvent is typically treated as a structureless medium of a given dielectric
constant $ \varepsilon $ (for water, $\varepsilon \sim 80$.) The solute itself
is treated as a low-$\varepsilon$ ``cavity'' \cite{jac95}.

Clearly, this approach fails to take into account many relevant details of the
solvent structure. The presence of long-ranged electrostatic fields induces
static correlations among the polar molecules that may vary considerably in
space depending on the molecular build-up of the solute. These can be taken
into account by performing Molecular Dynamics simulations, thus leaving the
continuum description at the expense of high computational cost even for small
molecules \cite{levy}.

Including more structural detail into the continuum description on the one
hand without increasing computational complexity on the other is therefore
highly desirable. At the same time, however, also the avoidance of additional
parameter sets, which is common practice in the use of continuum
electrostatics in computational chemistry, is a key requirement.

In this work, we follow one particular route to achieve this goal.  We develop
an approach for the computation of the electrostatic (polar) contribution to
the free energy of solvation based on nonlocal electrostatics
\cite{nonlocal,kornyshev1,kornyshev2,sutmann,basilevsky1,basilevsky2}.  This
continuum approach, originating in the physics literature as a generalization
of classical electrostatics {\it to account for media with spatial
dispersion}, is developed here within a physically transparent minimal
parameter set and applied to the simplest systems of single ions and small
molecules for which experimental and theoretical results are available for
quantitative comparison. In particular we address the issue of a systematic
choice of ion radii which most commonly is based on essentially empirical
parametrizations \cite{lee}. We conclude with an outlook on extensions of our
approach for more complex, and biologically relevant, molecules.

\section{Nonlocal electrostatics of the solvent}
\label{2}

The electrostatic potential $\phi$ of a charged molecule in solution is,
within classical (local) electrostatics, given by the Poisson equation
\cite{jackson}
\begin{equation} \label{one}
    \nabla\left[\varepsilon({\bf r})\varepsilon_0\nabla \phi({\bf r})\right] = -\varrho({\bf r})
\end{equation}
where $\varepsilon_0$ denotes the dielectric constant of vacuum,
and $\varepsilon({\bf r})$ is a local dielectric function which is
taken as the constants $ \varepsilon_{water} \approx 78 $ of the
solution (water), and a much smaller value within a solute
molecule (for proteins, e.g., a typical value taken is
$\varepsilon \approx 2-5 $). The rhs of eq.(\ref{one}) is the
charge density giving rise to the potential.

This arguably simple description is prone to complications: first,
the transition region between solute and solution is ill-described
and needs a further (rather delicate) modeling. Second, very
little is said about structural effects due to the orientation of
the polar water molecules near highly charged regions of a given
solute molecule.

The simplest way to introduce structural effects into a continuum
description of the solvent is to account for correlations due to
polarization effects between solution molecules characterized by a
correlation length $\lambda$. This gives rise to a nonlocal
generalization of eq.(\ref{one}) given by \cite{nonlocal}
\begin{equation} \label{two}
    \nabla_{\bf r}\int dV' \varepsilon(|{\bf r} - {\bf r'}|)\varepsilon_0\nabla_{\bf r'} \phi({\bf r'}) = -\varrho({\bf r})
\end{equation}
where isotropy of the liquid medium is supposed. Eq.(\ref{two})
needs to be accompanied by suitable boundary conditions (see
\cite{basilevsky1}). These conditions can be simplified
considerably by assuming the atomic constituents as conducting
spheres, which is exact for the case of ions, and approximate for
the case of small molecules. The solution of the nonlocal
solvation problem with a treatment of the full boundary conditions
is beyond the scope of this work and will be reported elsewhere
\cite{we}.

The rationale behind eq.(\ref{two}) is the following. Since the
electrostatic fields are long-ranged, the solvent molecules will
feel the presence of the fields of other molecules over
characteristic distances (that will be of particular relevance
below). Maintaining locality of the physical fields (i.e.,
electrostatic potentials and electric fields), it is the response
functions into which this purely static correlation effect needs
to be embedded. We note that the dielectric function is related to
the dielectric response function via $\varepsilon = 1 + \chi$,
irrespective of the local or nonlocal character of the theoretical
description.

Due to the assumption of spatial isotropy, eq.(\ref{two}) lends
itself to a treatment in Fourier space, reducing the linear
integro-differential equation eq.(\ref{two}) to an algebraic
equation involving the Fourier-transformed function
$\varepsilon(|{\bf k}|) $ with wavevector ${\bf k} $. Explicit
functional dependences of either $\varepsilon(|{\bf r} - {\bf
r'}|)$ or $ \varepsilon(|{\bf k}|) $ have been derived from
various approaches, e.g. within a Ginzburg-Landau theory for the
polarization fields \cite{sutmann}.

Motivated by these works we have formulated a family of nonlocal
dielectric functions fulfilling the following minimal requirements
\cite{anhi-diplom} :

i) for high $k$-values (equivalently, on small spatial scales) the
solvent molecules cannot follow the polarization forces.

Thus, a limiting value for the macroscopic dielectric function has
to be reached in this limit. Typical choices in the literature are
$\lim_{k\rightarrow\infty} \varepsilon(k) \sim
\varepsilon_{\infty} \approx 1.8 $ \cite{sutmann,hasted}, or
$\varepsilon_{\infty} = 1$ \cite{landau}. We have considered both
cases and do not find a significant dependence on this choice;
\\

ii) in the opposite limit, $k\rightarrow 0 $, the value for
$\varepsilon$ should equal the macroscopic value;
\\

iii) the physical length scale of the polarization fluctuations is
characterized by a correlation length $\lambda$. The local
electrostatic limit is given by $ lim_{\lambda \rightarrow 0}
\varepsilon(|{\bf r} - {\bf r'}|) = \varepsilon_{loc} \delta({\bf
r} - {\bf r'}) $;
\\

iv) causality conditions (Kramers-Kronig relations) need to be
fulfilled \cite{landau}.
\\

While the conditions i)-iv) are clearly not sufficient to
determine the dielectric function unambiguously, we find that the
following class of functions in Fourier space fulfill these
criteria sufficiently well \cite{fourier}
\begin{equation} \label{three}
    \varepsilon(k) = \frac{1}{(2\pi)^{3/2}}\left[\varepsilon_{\infty} -
\frac{\varepsilon_{\infty} - \varepsilon_{loc}}{(1 +
\lambda^2k^2)^n}\right]\, ,
\end{equation}
which depend on only two parameters, $ n $, and $ \lambda $. For $
n=2$, the dielectric function decays exponentially in real-space,
while $\varepsilon(k)$ is a Lorentzian for $n=1$, decaying in
real-space as $ \exp(-r/\lambda)/r$. Note that formally the case
$n=1$ was discussed previously in \cite{kornyshev1,kornyshev2},
based, however, on a different physical interpretation (see also
\cite{ritschel} for a similar approach in the context of nuclear
physics). We have also tested alternative choices, e.g. a Gaussian
model and a nonlocal model derived from a Ginzburg-Landau theory
for polarization fields \cite{sutmann}.

The comparison of our nonlocal models with other approaches and
experimental data is made possible by using the concept of an {\it
effective} local dielectric function $\widehat{\varepsilon}(r)$.
It is usually defined for the potential of a point charge $q$,
\begin{equation} \label{four}
    \phi(r) = \frac{1}{4\pi\varepsilon_0}\frac{q}{\widehat{\varepsilon}(r)r}
\end{equation}
where
\begin{equation} \label{five}
    \widehat{\varepsilon}(r)
    \equiv
    \left[\int
        \frac{d{\bf k}}{(2\pi)^{5/2}}
        \frac{\sin(kr)}{kr}
        \frac{1}{\varepsilon(k)}
    \right]^{\, -1}\, .
\end{equation}
For a general charge distribution $\widehat{\varepsilon}(r)$ can
be defined as well \cite{we}. We note that our determination of
the effective dielectric constant is consistently performed within
the nonlocal continuum theory and does not rely on data-based fits
\cite{meh84,mallik}.

Expression (\ref{five}) allows to test our approach against
previous results. We took empirical models for the radial
dependence of $\varepsilon$ derived for experimental results from
Mehler and Eichele (ME) \cite{meh84} and Conway (CO) \cite{con51}.
The CO-model is reproduced best by the Lorentzian ($n=1$) for
$\lambda = 15 $\AA, while the ME-model is reproduced best for
$\lambda = 24.13$\AA. The purely exponential model ($n=2$) displays
unphysical singularities in $\widehat{\varepsilon}(r)$ at small
$r$, while with a choice of $\lambda= 5$\AA, the CO-model is
reproduced for larger distances, $ r > 15$\AA. Figure 1 shows
$\widehat{\varepsilon}(r)$, as computed by eq.(5) for a single
sodium ion, using the Lorentzian model with $\lambda = 15$\AA. The
computation of $\widehat{\varepsilon}(r)$ performed here makes use
of a specifically adapted FFT applicable also to molecules
\cite{we}.

A Gaussian choice for $\varepsilon(k)$, however, leads to strong
oscillations in the effective dielectric function, although its
overall shape resembles the CO-model for $\lambda=5$\AA. Checking
our approach against the nonlocal model by Sutmann et al.
\cite{sutmann}, we find that their theory fails to comply with our
requirement iii), i.e., it does not reproduce the correct limiting
value at large distances.


\section{Single ions and the choice of their radii}
\label{3}

We now turn to the application of our approach to small ions. We
treat these ions as Born spheres. A standard problem in the
definition of Born-type ions is the definition of the ion radius
\cite{lee}. Starting from the accepted interpretation that a
solvated ion is surrounded by solvation shells, the first of these
shells will be ``as close as possible'' to the ion. We can
therefore identify the position of the centers of the oxygen in
the first solvation shell with the first peak in the radial
distribution function (rdf) which can be obtained either from
scattering experiments or from molecular dynamics simulations.

We defined the radius of the water oxygen as half the position of the
first peak in the oxygen-oxygen radial distribution function (rdf) of
bulk water. The ion radii were then derived by subtracting this radius
from the position of the first peak of the ion-oxygen rdf derived from
a molecular dynamics simulation.

As a specific input for our calculations we take the values for the
ion radii determined by {\AA}qvist \cite{Aqv90}, obtained from a
combined free energy perturbation/force field approach based on the GROMOS
force field \cite{vGb}, employing several different water models (flexible SPC,
rigid SPC, and TIP3P). Based on this input, we have determined the solvation
free energy from nonlocal electrostatics, which is given by
\begin{equation} \label{six}
    \Delta G^{polar} = -\frac{1}{2\varepsilon_0} \int d{\bf k} \left\{1 -
    \frac{1}{(2\pi)^{3/2}\varepsilon({\bf k})} \right\} D^2({\bf k})
\end{equation}
where ${\bf D}({\bf k}) $ is the dielectric flux density.

The result of the calculation is shown in Figs. 2 for monovalent
ions and in Fig. 3 for divalent ions, for all models to be
compared here. Note that we chose the parameter $\lambda$ based on
its fit to the effective dielectric function. The CO-model was
reproduced by $\lambda = 5${\AA} for $n=2$ (exponential model), by
$\lambda = 15${\AA} for $n=1$ (Lorentzian) and by $\lambda =
5${\AA} for the Gaussian model. The ME-model could only be
reproduced by $n=1$ with $\lambda = 24.13${\AA}. The value for the
exponential model is in accord with the findings in Ref.
\cite{basilevsky1}. Also note, as shown in Table 1, that the
values we obtain are only marginally corrected (if at all) by
effects due to nonlinear saturation, which contributes in
principle as well to a reduction of the dielectric constant near
the ion.

The model by Sutmann was used with two different limiting values
(Sutmann 1 with $\varepsilon_{\infty} = 1$, and Sutmann 2 with
$\varepsilon_{\infty} = 1.8$). As the figures show, our results
are consistently better than all the other theoretical curves.

The computation of electrostatic solvation free energies is also
possible by employing eq.(\ref{five}) in a standard (local)
Poisson solver. We have implemented this in an available library
(BALL) \cite{okhpl,we}. Fig. 4 compares the results of our
computations for the monovalent ions with the nonlocal theory
based on eq.(\ref{six}) and from the effective local dielectric
function, eq.(\ref{five}), demonstrating the consistency of both
approaches. Fig. 5 finally compares the theoretical results of the
local and nonlocal approaches, obtained with the Poisson solver,
with the experiment values. Evidently, the nonlocal approach
yields results consistently superior to those obtained from the
standard local theory.

We stress the significant advantage of our computations to work
with first-principle radii without {\it arbitrary} adjustments. In
Figs. 2-5, the {\AA}qvist-radii were used \cite{Aqv90}. In
addition, we have also tested the Shannon-radii derived from X-ray
crystal data \cite{sha1,sha2}, without significant effect on our
results. Our result thus give a basis to the general belief that
the need to introduce effective radii is {\it in fact} a
consequence of the local water structure around the ions
\cite{sandberg}. We believe that our nonlocal approach
demonstrates that it is therefore preferable conceptually to
introduce the length-scale governing the structural effects in the
solvent rather than introducing necessarily artificial procedures
to adjust the Born radii.

\section{Small molecules}
\label{4}

As a second application we apply our approach to determine the
solvation free energy for small alcohols for which we can assume that
the polar contribution exceeds the nonpolar one. Fortunately, accurate
measurements of the solvation free energies of these molecules are
available. We write the charge distribution of these molecules as a
linear superposition of radially symmetric partial distributions
translated by a vector ${\bf R}$,
$\varrho_i = \varrho(|{\bf r} + {\bf R}_i|) $, i.e. $\varrho({\bf r})
= \sum_{i=1}^{N} \varrho_i(|{\bf r} + {\bf R}_i|) $. Its Fourier
transform is then given by the expression $ \varrho({\bf k}) =
\sum_{i=1}^{N} \varrho_i(k) \exp(-i{\bf R_i}\cdot {\bf k}). $

Again we have to define the radii of the atoms. For this we chose to classify
the atoms into classes depending on their chemical environment in the molecule
(e.g., like the hydrogen atom in an OH-group), expecting that the radii of all
atoms in a certain class are more or less similar. For methanol ($\mathrm
CH_3OH$), e.g., we defined four classes of atoms and used their rdf's with the
oxygen of water to find the following set of radii: Methyl C: 2.135 {\AA},
hydroxyl O: 2.014 {\AA}, hydroxyl H: 1.115 {\AA}, methyl H:  1.394 {\AA}.

Compared to the calculation of the free energies for the ions this
computation is slightly more involved as it requires a
three-dimensional integration instead of a one-dimensional one.
Within the framework of the simplified boundary conditions, the
unperturbed fields of the atomic constituents are simply
superimposed. To ensure short computation times, in our
implementation we have used the VEGAS-Monte-Carlo integration
scheme \cite{VEGAS} that is supplied with the GNU scientific
library GSL \cite{gsl}. Details of these computations will again
be given elsewhere \cite{we}.

The results for some small alcohols are shown in Table 1. For the
comparison we have used the exponential with $\lambda=5$ \AA, the
Lorentzian with $\lambda=15$ \AA, and the Lorentzian with
$\lambda=24.13$ \AA. All values in these tables are given in
kJ/mol.

The interpretation of the results is more complicated than in the
Born ion case. While there we could assume that the contribution
of the nonpolar part of the free energy of solvation could be
neglected (for the Born ions it can be estimated to be of the
order of $10-20$ kJ/mol), this is not the case for the alcohols.
The electrostatic contribution is still the dominant part, but not
the only significant one. We therefore applied a very simple model
for the nonpolar contribution \cite{uhlig} in order to be able to
compare our results to experimental data. The results given for
the polar contribution can also be optimized by considering
improved charge distributions \cite{we}.

\section{Conclusions and outlook}
\label{5} We have demonstrated that the approach to the
computation of solvation free energies based on nonlocal
electrostatics is able to reproduce experimental data for ions and
small molecules with reasonable accuracy. The approach we put
forward has the basic advantage to rely on essentially one
parameter which has a transparent physical interpretation as the
correlation length of the polarization fluctuations. While within
the present paper this parameter is used as a fitting parameter,
it should be clear that it might be also determined by experiment
or simulation. A further significant advantage is that our
approach does not make use of the commonly used adjustments of
atomic or ionic radii. A challenge for the future will be the
extension of our approach to determine electrostatic solvation
free energies for more complex and biologically relevant
molecules. Work in this direction is in progress \cite{we}.
\\

{\bf Acknowledgement.} This work is supported by the DFG under
Schwerpunktsprogramm ``Informatikmethoden zur Analyse und
Interpretation grosser geno\-mischer Datenmengen" (grant
LE952/2-1).

\newpage

\begin{figure}[htpb]
  \begin{center}
    \epsfig{figure=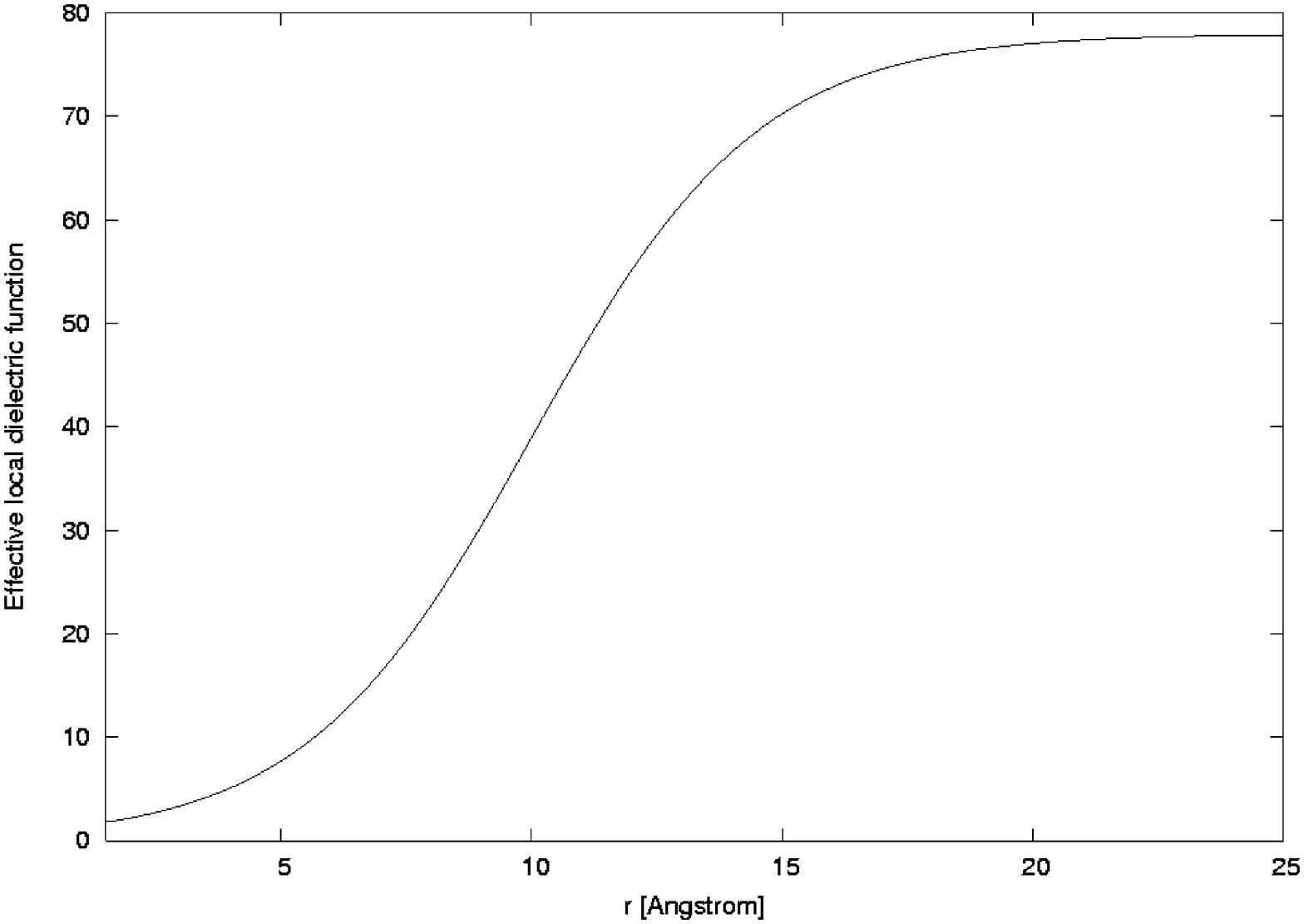,width=10cm}
    \caption{$\widehat{\varepsilon}(r)$ for a single sodium ion}
  \end{center}
\end{figure}
\newpage

\begin{figure}[htpb]
  \begin{center}
    \epsfig{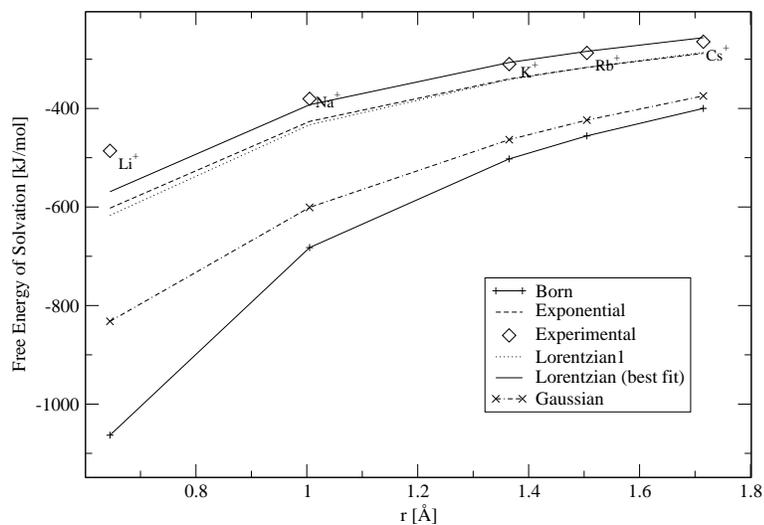}
    \caption{Free energy of solvation for monovalent ions}
    \end{center}
\end{figure}
\vspace{1.5cm}

\begin{figure}[htpb]
  \begin{center}
    \epsfig{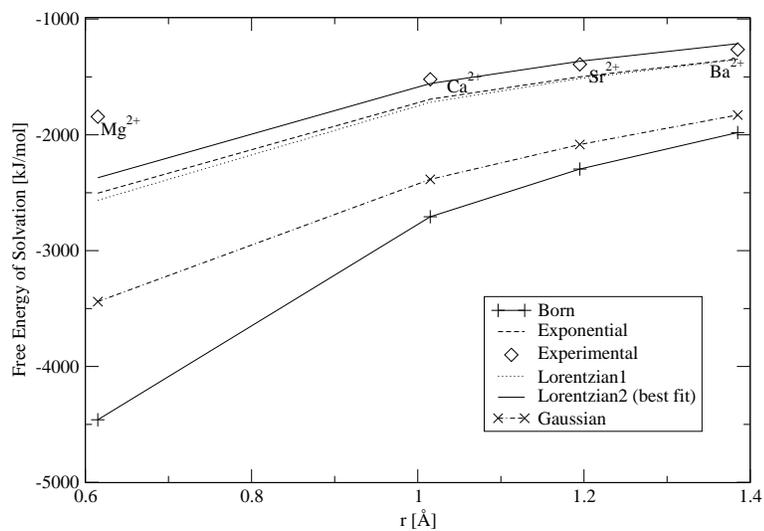}
    \caption{Free energy of solvation for divalent ions}
    \end{center}
\end{figure}

\begin{figure}[htpb]
  \begin{center}
    \epsfig{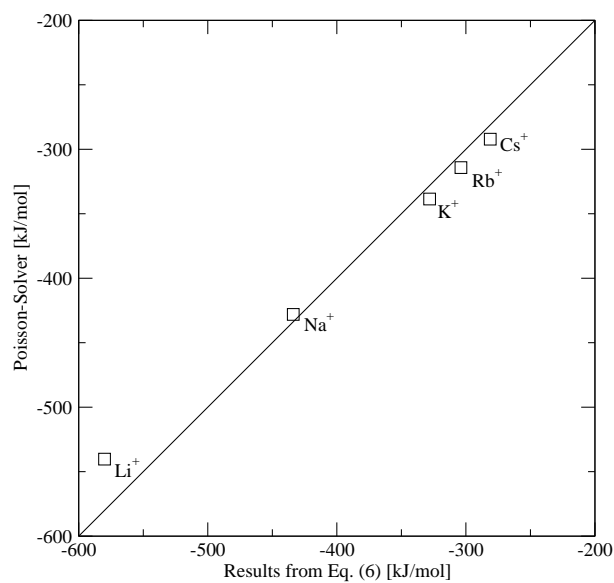}
    \caption{Free energy of solvation for monovalent ions, from the nonlocal
        expression, eq.(\ref{six}),
    and from the solution of the local Poisson equation with the effective dielectric constant, eq.(\ref{five}).}
    \end{center}
\end{figure}
  \begin{center}

\begin{figure}[htpb]
    \epsfig{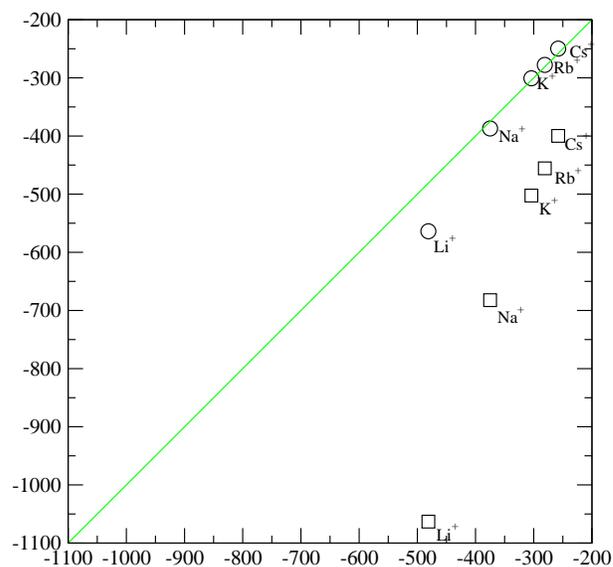}
    \caption{Free energy of solvation for monovalent ions, from the nonlocal
            expression, eq.(\ref{six}),
    (circles) and from the solution of the local Poisson equation (squares), i.e. the
        Born energies. (The experimental values corresponding to the ions are on the diagonal.)}
\end{figure}
\begin{table}
    {\scriptsize
    \begin{tabular}{|cc|cccc|cc|c|}
                \hline
        Ion & r [$\rm\AA$]
                & $\Delta G_{B}$
                & $\Delta G_{nloc}$
                & $\Delta G_{npol}$
                &   $\Delta\Delta G_{corr}$
                & $\Delta G_{calc}$
                & $\Delta G_{calc}^{corr}$
                & $\Delta G_{exp}$
        \\\hline
            $\mathrm{Li^+}$    & 0.645 & -1063.08 & -568.9 & 5.03 & 0   & -563.87 & -563.87& -481 \\
            $\mathrm{Na^+}$    & 1.005 & -682.30  & -392.51 & 5.48 & 0   & -387.03 & -387.03& -375 \\
            $\mathrm{K^+}$     & 1.365 &  -502.36 & -306.79 & 6.01 & 0  & -300.78 & -300.78& -304 \\
            $\mathrm{Rb^+}$    & 1.505 &  -455.63 & -284.02 & 6.23 & 0  & -277.79 & -277.79& -281 \\
            $\mathrm{Cs^+}$    & 1.715 &  -399.84 & -256.39 & 6.59 & 0  & -249.8 & -249.8& -258 \\
                        \hline
            $\mathrm{Mg^{2+}}$ & 0.615 & -4459.74 &-2370.66 & 4.99 & 110& -2365.67 & -2255.67&-1838\\
            $\mathrm{Ca^{2+}}$ & 1.015 & -2707.32 &-1557.41 & 5.49 & 45 & -1551.92 & -1506.92&-1515 \\
            $\mathrm{Sr^{2+}}$ & 1.195 & -2295.29 &-1364.6 & 5.75 & 30  & -1358.85 & -1328.85&-1386 \\
            $\mathrm{Ba^{2+}}$ & 1.385 & -1980.43 &-1213.12 & 6.04 & 24 & -1207.08 & -1183.08&-1259 \\
            \hline
        \end{tabular}
    }
\vspace{1cm}
           \caption{ Comparison of different models for the hydration free
            energy of different mono- and divalent cations. All energies are in kJ/mol.
                        $\Delta G_{B}$ is the hydration free energy computed for a Born ion of
            radius $r$.
            $\Delta G_{nloc}$ is the
            electrostatic hydration free energy computed with our nonlocal Lorentzian
            model ($\lambda = 24.13$).
            $\Delta G_{npol}$ is the nonpolar
            contribution to the hydration free energy as computed with the
            model of Uhlig \cite{uhlig}.
            $\Delta G_{corr}$ is the nonlinear
            correction to the hydration free energy for the corresponding
            ion (taken from \cite{sandberg}).
            $\Delta G_{exp}$ is
            the experimental hydration free energy (taken from~\cite{exp}).
            $\Delta G_{calc} = \Delta G_{nloc} + \Delta G_{npol}$.
            $\Delta G_{calc}^{corr} = \Delta G_{calc} + \Delta\Delta G_{corr}$. }
\end{table}

\begin{table}[hp]
  \centering
    \begin{tabular}[c]{|c|c|c|c|c|c|}
      \hline
      Substance & Exponential &
      Nonpolar & $\sum^{ }$ & Experiment \\
      & $\lambda=5$ \AA & & & \\
      \hline
      Ethanol & -32.84 & 8.89 & -23.95 & -20.51\\
      \hline
      Methanol & -25.64 & 8.10 & -17.54 & -21.26\\
      \hline
      1-Butanol & -29.79 & 10.31 & -19.48 & -19.76\\
      \hline
      1-Hexanol & -37.41 & 11.83 & -25.58 & -18.25\\
      \hline
      Octanol & -33.63 & 13.28 & -20.35 & -17.12\\
      \hline
      Cyclopentanol & -28.90 & 10.31 & -18.59 & -22.98\\
      \hline\hline
      & Lorentzian & & &\\
      & $\lambda=15$ \AA & & &\\
      \hline
      Ethanol & -34.17 & 8.89 & -25.28 & -20.51\\
      \hline
      Methanol & -29.49 & 8.10 & -21.39 & -21.26\\
      \hline
      1-Butanol & -31.13 & 10.31 & -20.82 & -19.76\\
      \hline
      1-Hexanol & -38.88 & 11.83 & -27.05 & -18.25\\
      \hline
      Octanol   & -35.08 & 13.28 & -21.79 & -17.12\\
      \hline
      Cyclopentanol & -30.16 & 10.31 & -19.85 & -22.98\\
      \hline
      \hline
       & Lorentzian &
       & &  \\
      & $\lambda=24.13$ \AA & & &\\
      \hline
      Ethanol & -32.89 & 8.89 & -24.0 & -20.51\\
      \hline
      Methanol & -27.88 & 8.10 & -19.78 & -21.26\\
      \hline
      1-Butanol & -29.94 & 10.31 & -19.63 & -19.76\\
      \hline
      1-Hexanol & -37.65 & 11.83 & -25.82 & -18.25\\
      \hline
      Octanol & -33.77 & 13.28 & -20.49 & -17.12\\
      \hline
      Cyclopentanol & -29.06 & 10.31 & -18.75 & -22.98\\
      \hline
    \end{tabular}
    \caption{Results for the free energy of solvation for small
        alcohols. All free energies are given in kJ/mol.}
    \label{fig:alcres}
\end{table}

\end{center}

\begin{thebibliography}{}
\bibitem{fourier} We have normalized the Fourier transform symmetrically.
\end{thebibliography}
\end{document}